\documentstyle[12pt,epsf]{article}
\begin{document}

\title{Simple predictions from ALCOR$_c$\\
for rehadronisation of charmed quark matter
  } 
\author{ P. L\'evai, T.S. Bir\'o, T. Cs\"org\H o, J. Zim\'anyi\\[1ex]
RMKI Research Institute for Particle and Nuclear Physics, \\
\ \  P. O. Box 49, Budapest, 1525, Hungary }

\date{July  22.      2000}

\maketitle
\begin{abstract}
We study the production of  charmed hadrons with the help of ALCOR$_c$,
the algebraic coalescence model for rehadronisation of charmed
quark matter.
Mesonic ratios are introduced as factors connecting various
antibaryon to baryon ratios. 
The resulting simple relations 
could serve as tests of quark matter formation and
coalescence type rehadronization in heavy ion collisions.
\end{abstract}

Charm hadron production has gained an enhanced attention in relativistic heavy
ion collisions at CERN SPS (Super Proton Synchrotron).
The measured  anomalous suppression of the $J/\psi$  
in Pb+Pb collision \cite{NA50supr} is considered as one of the
strongest candidates for an 
evidence of quark-gluon plasma (QGP) formation in Pb+Pb collision
at 158 GeV/nucleon bombarding energy \cite{QGPjpsi}. 
So far only the $J/\psi$ and $\psi'$ production was measured 
in heavy ion collisions through dilepton
decay channels. However, recent efforts to measure D-meson
\cite{letterNA6i,NA49D} support the theoretical investigation
of charm production from a different point of view. Namely,
it is interesting to search for predictions on the
total numbers of charmed hadrons and their ratios. 
The answer to this question may become
very  important at the RHIC accelerator, where a large
number of charmed quark-antiquark pairs will be produced and a
number of different charmed hadrons could be detected. 

In this paper we assume that quark matter is formed in heavy ion
collisions and the charm hadrons will be produced directly from
this state via quark coalescence.
Quark coalescence was successfully applied to describe direct hadron
production from deconfined quark matter phase 
(see. the ALCOR \cite{ALCOR1,ALCOR3},
the Transchemistry \cite{Transchem} and the MICOR 
\cite{MICOR} models).
In these models the hadronic rescatterings are assumed to be weak and
they are neglected. Thus the results of quark coalescence processes
were compared directly to the experimental data - and the agreement was
remarkably good.

Multicharm hadron production was already investigated in a simplified
quark coalescence model  and first results were obtained
at RHIC and LHC energies, where an appreciable number of charm quark may
appear \cite{multic}.
Here we summarize simple predictions from a non-linear
algebraic coalescence model ALCOR$_c$, the extension
of the ALCOR model of algebraic coalescence of strange quark matter
~\cite{ALCOR1,ALCOR3} for the inclusion of charmed quarks, mesons and 
baryons.

The description of the charmed baryons 
has to deal with the fact that two possible
$(1/2)^+$ baryon multiplets 
exist containing $c$, $s$ and $u$ (or $d$) quarks, one 
being flavor symmetric under $s$ and $d$ (or $u$) exchange and the other 
being antisymmetric \cite{csym}.  
The heavier (symmetric) states decay into the lighter (antisymmetric) one
by emission of a $\gamma$ or a $\pi$ meson. 
However, if quark clusterization is the basic hadronization 
process, then the effect of
these decay processes will be cancelled from charmed antibaryon
to baryon ratios.
Neglecting the difference between the light $u$ and $d$ quarks and using the
notation $q$ for them, 
the 10 different types of produced quark clusters can be connected 
to the 40 lowest lying SU(4)-flavor baryon species in the following way
(see e.g. Ref.~\cite{Charmb,PDG} for precise quark content, 
hadron names and masses):
\begin{eqnarray}
N(qqq) &:=& \ p, \ n, \ \Delta^{++}, \ \Delta^+, \ \Delta^0, 
            \ \Delta^- \ ; \nonumber \\
Y(qqs) &:=& \ \Lambda^0, \ \Sigma^+, \ \Sigma^0, \ \Sigma^-, \ \Sigma^{*+},
            \ \Sigma^{*0}, \ \Sigma^{*-} \ ; \nonumber \\
\Xi(qss) &:=& \ \Xi^0, \ \Xi^-, \ \Xi^{*0}, \ \Xi^{*-} \ ; \nonumber \\
\Omega(sss) &:=& \ \Omega^- \ ; \nonumber \\
Y_c(qqc) &:=& \ \Lambda_c^+, \ \Sigma_c^{++}, \ \Sigma_c^+, \ \Sigma_c^0, 
              \ \Sigma_c^{*++}, \ \Sigma_c^{*+}, \ \Sigma_c^{*0} \ ;\nonumber \\
\Xi_c(qsc) &:=& \ \Xi_c^+, \ \Xi_c^0, \ {\Xi_c}'{^+}, \ {\Xi_c}'{^0},
                \ \Xi_c^{*+}, \ \Xi_c^{*0} \ ; \nonumber \\
\Omega_c(ssc) &:=& \ \Omega_c^0, \ \Omega_c^{*0} \ ; \nonumber \\
\Xi_{cc}(qcc) &:=& \ \Xi_{cc}^{+}, \ \Xi_{cc}^{++}, 
                \ \Xi_{cc}^{*+}, \ \Xi_{cc}^{*++} \ ; \nonumber \\
\Omega_{cc}(scc) &:=& \ \Omega_{cc}^+, \ \Omega_{cc}^{*+} \ ; \nonumber \\
\Omega_{ccc}(ccc) &:=& \ \Omega_{ccc}^{++} 
\end{eqnarray}
 
In ALCOR, the algebraic coalescence model of rehadronization it is assumed that
the number of directly produced hadrons is given by the product of the 
the number of quarks (or anti-quarks) from which those hadrons are
produced, multiplied by coalescence coefficients $C_h$ and 
by non-linear normalization coefficients $b_q$, that take into account
conservation of quark numbers during quark coalescence, as will be explained
subsequently.
The number of various hadrons and quarks is denoted by the 
symbol usual for that type of particles, e.q. $q$, $s$ and $c$ 
denote the number of light, strange and charmed quarks, respectively,
$N$ denotes the number of protons, neutrons and deltas etc.

In this way the
baryons and antibaryons 
can be described through the following clustering relations:
\begin{eqnarray}
N\,(qqq)&=&
C_N \cdot (b_q\, q)^3  \hskip 3.28 truecm 
{\overline N}\, ({\overline q} {\overline q} {\overline q})
 = C_{\overline N} \cdot (b_{\overline q} \, {\overline q})^3
\nonumber \\
Y\,(qqs) &= &
C_{Y} \cdot (b_q\, q)^2 \cdot (b_s\, s) \hskip 2.15 truecm
{\overline Y} \, ({\overline q} {\overline q} {\overline s}) = 
C_{\overline Y} \cdot (b_{\overline q}\, {\overline q})^2 
     \cdot (b_{\overline s}\, {\overline s}) \nonumber \\
\Xi\,(qss) &= &
C_{\Xi} \cdot (b_q\, q) \cdot (b_s\, s)^2 \hskip 2.22 truecm
{\overline \Xi}\,({\overline q} {\overline s} {\overline s}) = 
C_{\overline \Xi} \cdot (b_{\overline q}\, {\overline q}) 
     \cdot (b_{\overline s} \, {\overline s})^2 \nonumber \\
\Omega\,(sss) &= &
C_{\Omega} \cdot (b_s\, s)^3 \hskip 3.43 truecm
{\overline \Omega} \,({\overline s} {\overline s} {\overline s}) = 
C_{\overline \Omega} 
     \cdot (b_{\overline s}\, {\overline s})^3 \nonumber \\ 
Y_c\,(qqc) &= &
C_{Y}^c \cdot (b_q\, q)^2 \cdot (b_c\, c) \hskip 2.1 truecm
{\overline Y}_c \, ({\overline q} {\overline q} {\overline c}) = 
C_{\overline Y}^c \cdot (b_{\overline q}\, {\overline q})^2 
     \cdot (b_{\overline c}\, {\overline c}) \nonumber \\
\Xi_c\,(qsc) &= &
C_{\Xi}^c \cdot (b_q\, q) \cdot (b_s\, s) \cdot (b_c\, c) \hskip 1.1 truecm
{\overline \Xi}_c \, ({\overline q} {\overline s} {\overline c}) = 
C_{\overline \Xi}^c \cdot (b_{\overline q}\, {\overline q})
     \cdot (b_{\overline s}\, {\overline s})
     \cdot (b_{\overline c}\, {\overline c}) \nonumber \\
\Omega_c\,(ssc) &= &
C_{\Omega}^c \cdot (b_s\, s)^2 \cdot (b_c\, c) \hskip 2.1 truecm
{\overline \Omega}_c \, ({\overline s} {\overline s} {\overline c}) = 
C_{\overline \Omega}^c \cdot (b_{\overline s}\, {\overline s})^2 
     \cdot (b_{\overline c}\, {\overline c}) \nonumber \\
\Xi_{cc}\,(qcc) &= &
C_{\Xi}^{cc} \cdot (b_q\, q) \cdot (b_c\, c)^2 \hskip 2 truecm
{\overline \Xi}_{cc}\,({\overline q} {\overline c} {\overline c}) = 
C_{\overline \Xi}^{cc} \cdot (b_{\overline q}\, {\overline q}) 
     \cdot (b_{\overline c} \, {\overline c})^2 \nonumber \\
\Omega_{cc}\,(scc) &= &
C_{\Omega}^{cc} \cdot (b_s\, s) \cdot (b_c\, c)^2 \hskip 2 truecm
{\overline \Omega}_{cc}\,({\overline s} {\overline c} {\overline c}) = 
C_{\overline \Omega}^{cc} \cdot (b_{\overline s}\, {\overline s}) 
     \cdot (b_{\overline c} \, {\overline c})^2 \nonumber \\
\Omega_{ccc} \,(ccc) &= &
C_{\Omega}^{ccc} \cdot (b_c\, c)^3 \hskip 3.1 truecm
{\overline \Omega}_{ccc} \,({\overline c} {\overline c} {\overline c}) = 
C_{\overline \Omega}^{ccc} 
     \cdot (b_{\overline c}\, {\overline c})^3 
\label{cbarion}
\end{eqnarray}

\newpage
Mesons in the pseudoscalar and vector SU(4)-flavor multiplets are grouped
in the following way:
\begin{eqnarray}
\pi(q{\overline q}) &:=& \ \pi^+, \ \pi^0, \ \pi^-, \ \eta, 
            \ \rho^+, \ \rho^0, \ \rho^-, \ \omega;  \nonumber \\
K(q{\overline s}) &:=& \ K^+, \ K^0, 
            \ K^{*+}, \ K^{*0}; \nonumber \\
{\overline K}({\overline q}s) &:=& \ K^-, \ {\overline K}^0, 
            \ K^{*-}, \ {\overline K}^{*0}; \nonumber \\
\phi(s{\overline s}) &:=& \ {\eta}', \ \phi; \nonumber \\
D({\overline q}c) &:=& \ D^+, \ D^0, 
            \ D^{*+}, \  D^{*0}; \nonumber \\
{\overline D}(q{\overline c}) &:=& \ D^-, \ {\overline D}^0, 
            \ D^{*-}, \  {\overline D}^{*0}; \nonumber \\
D_s({\overline s}c) &:=& \ D_s^+,  \ D_s^{*+}; \nonumber \\
{\overline D}_s(s{\overline c}) &:=& \ D_s^-,  \ D_s^{*-}; \nonumber \\
J/\psi({\overline c}c) &:=& \ \eta_c, \ J/\psi; 
\end{eqnarray}

Thus the number of directly produced mesons reads as

\begin{eqnarray} 
\pi \, (q\overline q) &=& 
C_\pi \cdot (b_q \, q) \cdot (b_{\overline q} \, {\overline q}) 
    \hskip 2.1 truecm
J/\psi \, (c\overline c) = 
C_{J/\psi} \cdot (b_c \, c) \cdot (b_{\overline c} \, {\overline c}) 
   \nonumber \\
K \, (q \overline s) &=& 
C_K \cdot (b_q \, q) \cdot (b_{\overline s} \, {\overline s}) 
    \hskip 2.4 truecm
D \, ( {\overline q} c) = 
C_D \cdot  (b_{\overline q} \, {\overline q}) \cdot (b_c \, c) \nonumber \\
{\overline K} \, (\overline q s) &=& 
C_{\overline K}
\cdot (b_{\overline q} \, {\overline q}) \cdot (b_s \, s) 
    \hskip 2.4 truecm
{\overline D} \, (q \overline c ) = 
C_{\overline D}
\cdot (b_q \, q) \cdot (b_{\overline c} \, {\overline c})  \nonumber \\
\phi \, (s \overline s) &=& 
C_{\phi} 
\cdot (b_s \, s) \cdot (b_{\overline s} \, {\overline s}) 
    \hskip 2.4 truecm
D_s \, ( \overline s c) =  
C_D^s \cdot (b_{\overline s} \, {\overline s}) \cdot (b_c \, c) \nonumber \\
  &\ &  \hskip 5.3 truecm
{\overline D}_s \, (s \overline c) =  
C_{\overline D}^s
\cdot (b_s \, s) \cdot (b_{\overline c} \, {\overline c})  
\label{cmeson}
\end{eqnarray}

As a straightforward extension to the ALCOR model, 
the non-linear coalescence factors
$b_q$, $b_s$, $b_c$ and the 
$b_{\overline q}$, $b_{\overline s}$, $b_{\overline c}$  
are determined unambiguously from the requirement
that the number of the constituent quarks and anti-quarks do not  change
during the hadronization, and that all initially available 
quarks and anti-quarks
have to end up in the directly produced hadrons. 
This constraint is a basic assumption in all models of quark coalescence. 
The correct quark counting yields to the following
equations, expressing the conservation of the number of quarks: 
\begin{eqnarray}
q &=& 3 \ N\,(qqq) + 2\  Y\,(qqs) + \Xi\,(qss) 
      + K \, (q \overline s ) + \pi  \, (q \overline q) +
       \nonumber\\
  &&+ 2 \ Y_{c}\,(qqc)+ \Xi_c\,(qsc) + \Xi_{cc}\,(qcc)
+ {\overline D} \, (q\overline  c)  \\
{\overline q} &=& 
   3 \ {\overline N}\, 
       ({\overline q}{\overline q}{\overline q})
 + 2 \ {\overline Y}\,
       ({\overline q}{\overline q}{\overline s}) 
 + \overline{\Xi}\,({\overline q}{\overline s}{\overline s}) 
      +  {\overline K} \, (\overline q s ) + \pi \, (q \overline q) +
       \nonumber\\
  &&+ 2 \ {\overline Y}_{c}\,({\overline q}{\overline q}{\overline c})
    + {\overline \Xi}_c\,
       ({\overline q}{\overline s}{\overline c}) 
    + {\overline \Xi}_{cc}\,
       ({\overline q}{\overline c}{\overline c})
+  D\, (\overline q c) \\ 
s &=& 3\ \Omega \,(sss) + 2\ \Xi \,(qss) + Y\,(qqs) 
      + {\overline K} \, (\overline q s) + \phi  \, (s \overline s) +
       \nonumber\\
  &&+ 2 \ \Omega_{c}\,(ssc)+ \Xi_c\,(qsc) + \Omega_{cc}\,(scc)
+ {\overline D}_s \, (s\overline  c)  \\
{\overline s} &=& 
   3 \ {\overline \Omega}\, 
       ({\overline s}{\overline s}{\overline s})
 + 2 \ {\overline \Xi}\,
       ({\overline q}{\overline s}{\overline s}) 
 + \overline{Y}\,({\overline q}{\overline q}{\overline s}) 
      +   K \, (q \overline s ) + \phi \, (s \overline s) +
       \nonumber\\
  &&+ 2 \ {\overline \Omega}_{c}\,({\overline s}{\overline s}{\overline c})
    + {\overline \Xi}_c\,
       ({\overline q}{\overline s}{\overline c}) 
    + {\overline \Omega}_{cc}\,
       ({\overline s}{\overline c}{\overline c})
+  D_s\, (\overline s c) \\ 
c &=& 3 \ \Omega_{ccc}\,(ccc) + 2\  \Xi_{cc}\,(qcc) + \Lambda_c\,(qqc) 
      +  D \, (\overline q c) + J/\psi \, (c \overline c) +
       \nonumber\\
  &&+ 2 \ \Xi_{cc}\,(scc)+ \Lambda_c\,(qsc) + \Lambda_c\,(ssc)
+ D_s \, (\overline s c)  \\
{\overline c} &=& 
   3 \ {\overline \Omega}_{ccc}\, 
       ({\overline c}{\overline c}{\overline c})
 + 2 \ {\overline \Xi}_{cc}\,
       ({\overline q}{\overline c}{\overline c}) 
 + \overline{\Lambda}_c\,({\overline q}{\overline q}{\overline c}) 
      +  {\overline D} \, (q \overline c ) + J/\psi \, (c \overline c) +
       \nonumber\\
  &&+ 2 \ {\overline \Xi}_{cc}\,({\overline s}{\overline c}{\overline c})
    + {\overline \Lambda}_c\,
       ({\overline q}{\overline s}{\overline c}) 
    + {\overline \Lambda}_c\,
       ({\overline s}{\overline s}{\overline c})
+ {\overline D}_s \, (s \overline c)  
\end{eqnarray}

These equations for 
$q$, $s$, $c$ and
 $ ({\overline q}$, ${\overline s}$, ${\overline c}) $
 determine the six $b_i$ normalization factors --- which are not
free parameters. These constraints, together with the prescription
of the coalescence factors $C_i$, complete the description of
hadron production from charmed quark matter by quark coalescence,
and define the ALCOR$_c$ model.

In this paper, we will evaluate only the simplest predictions from
ALCOR$_c$, by considering ratios of the number of particles to the number
of anti-particles and by assuming the symmetry of the coalescence process for
charge conjugation, extending the results of ref.~\cite{ALCOR3} to the 
case of charmed quarks, mesons and baryons.

Assuming that the coalescence coefficients $C$ for hadrons are equal
to that for the corresponding anti-particles, e.g. 
$C_\Lambda = C_{\overline \Lambda}$, the following relations were obtained
for the ratio of light and strange antibaryons and baryons \cite{ALCOR3}: 

\begin{eqnarray}
\frac{{\overline N}\, ({\overline q} {\overline q} {\overline q})}
{N\,(qqq)} &=& 
\left[ \frac{b_{\overline q} \, {\overline q}}{b_q\, q} \right]^3 
\label{rn} \\
& & \nonumber \\
\frac{{\overline Y} \, ({\overline q} {\overline q} {\overline s})}
{Y\,(qqs)} &=& 
\left[ \frac{b_{\overline q} \, {\overline q}}{b_q\, q} \right]^2 \cdot 
\left[ \frac{b_{\overline s} \, {\overline s}}{b_s\, s} \right]  
\label{rlam}\\
& & \nonumber \\
\frac{{\overline \Xi} \, ({\overline q} {\overline s} {\overline s})}
{\Xi\,(qss)} &=& 
\left[ \frac{b_{\overline q} \, {\overline q}}{b_q\, q} \right] \cdot 
\left[ \frac{b_{\overline s} \, {\overline s}}{b_s\, s} \right]^2  
\label{rxi} \\
& & \nonumber \\
\frac{{\overline \Omega} \, ({\overline s} {\overline s} {\overline s})}
{\Omega\,(sss)} &=& 
\left[ \frac{b_{\overline s} \, {\overline s}}{b_s\, s} \right]^3  
\label{romega}
\end{eqnarray}

Inspecting eqs.~(\ref{rn})-(\ref{romega}) one can recognize, 
that the kaon to anti-kaon  ratio ${\cal S}^{qs}$ has a special role
as it acts as a stepping factor that connects various  antibaryon to baryon
rations,
\begin{equation}
{\cal S}^{qs} \equiv
\frac{K\,(q\overline s)}{{\overline K}\, ({\overline q} s )}=
\left[ \frac{b_q\, q}{b_{\overline q} \, {\overline q}} \right] \cdot
\left[ \frac{b_{\overline s} \, {\overline s}}{b_s\, s} \right] \ \ .
\label{oper_s}
\end{equation}

This factor ${\cal S}^{qs}$  substitutes a light quark with 
a strange quark in the antibaryon to baryon ratios.
Thus it  shifts the antibaryon to baryon
ratios and changes their strangeness content by one unit, 
as the following relations display:
\begin{eqnarray}
{\cal S}^{qs}  \left[ \frac{{\overline N}}{N} \right] &=&
\frac{{\overline Y}} {Y} \\
{\cal S}^{qs}  {\cal S}^{qs} \left[ \frac{{\overline N}}{N} \right] &=&
\frac{{\overline \Xi}} {\Xi} \\
{\cal S}^{qs} {\cal S}^{qs} {\cal S}^{qs}  
\left[ \frac{{\overline N}}{N} \right] &=&
\frac{{\overline \Omega}} {\Omega} 
\end{eqnarray}
 
 The inverse factor,
 ${\cal S}^{sq} = ({\cal S}^{qs})^{-1}$ decreases the strangeness
 content and increases the number of light quarks in the 
 antibaryon to baryon ratios.
 Note that these relations hold between the ratios of the
 directly produced anti-baryons to baryons and that the 
 number of observed anti-baryons and baryons have to be
 corrected to the various chains of resonance decays~\cite{ALCOR3}. 

\newpage

Extending the above ALCOR model to the case of
charmed baryons and antibaryons, further relations are obtained: 

\begin{eqnarray}
\frac{{\overline Y}_c \, ({\overline q} {\overline q} {\overline c})}
{Y_c\,(qqc)} &=& 
\left[ \frac{b_{\overline q} \, {\overline q}}{b_q\, q} \right]^2 \cdot 
\left[ \frac{b_{\overline c} \, {\overline c}}{b_c\, c} \right]  
\hskip 1.5 truecm
\frac{{\overline \Xi}_c \, ({\overline q} {\overline s} {\overline c})}
{\Xi_c\,(qsc)} = 
\left[ \frac{b_{\overline q} \, {\overline q}}{b_q\, q} \right] \cdot 
\left[ \frac{b_{\overline s} \, {\overline s}}{b_s\, s} \right] \cdot 
\left[ \frac{b_{\overline c} \, {\overline c}}{b_c\, c} \right]  
\nonumber \\
& & \nonumber \\
\frac{{\overline \Xi}_{cc} \, ({\overline q} {\overline c} {\overline c})}
{\Xi_{cc}\,(qcc)} &=& 
\left[ \frac{b_{\overline q} \, {\overline q}}{b_q\, q} \right] \cdot 
\left[ \frac{b_{\overline c} \, {\overline c}}{b_c\, c} \right]^2  
\hskip 1.5 truecm
\frac{{\overline \Omega}_{c} \, ({\overline s} {\overline s} {\overline c})}
{\Omega_{c}\,(ssc)} = 
\left[ \frac{b_{\overline s} \, {\overline s}}{b_s\, s} \right]^2 \cdot 
\left[ \frac{b_{\overline c} \, {\overline c}}{b_c\, c} \right]  
\nonumber \\
& & \nonumber \\
\frac{{\overline \Omega}_{ccc} \, ({\overline c} {\overline c} {\overline c})}
{\Omega_{ccc}\,(ccc)} &=& 
\left[ \frac{b_{\overline c} \, {\overline c}}{b_c\, c} \right]^3  
\hskip 2.9 truecm
\frac{{\overline \Omega}_{cc} \, ({\overline s} {\overline c} {\overline c})}
{\Omega_{cc}\,(scc)} =
\left[ \frac{b_{\overline s} \, {\overline s}}{b_s\, s} \right] \cdot
\left[ \frac{b_{\overline c} \, {\overline c}}{b_c\, c} \right]^2
\label{charm}
\end{eqnarray}

These ratios and the ratios from eqs.~(\ref{rn})-(\ref{romega})
can be organized into a special structure displayed in Fig.1.
We can introduce two more factors ${\cal S}^{sc}$ and
${\cal S}^{cq}$ constructed 
as in eq.(\ref{oper_s}) but from the ratios of charmed mesons:
\begin{eqnarray}
{\cal S}^{sc}  \equiv \ 
\frac{{\overline D}_s\,(s\overline c)}{ D_s\, ({\overline s} c )} &=&
\left[ \frac{b_s\, s}{b_{\overline s} \, {\overline s}} \right] \cdot
\left[ \frac{b_{\overline c} \, {\overline c}}{b_c\, c} \right] 
\label{oper_c} \\
{\cal S}^{cq}  \equiv \ 
\frac{ D\, ({\overline q} c )}{{\overline D}\,(q\overline c)} &=&
\left[ \frac{b_c\, c}{b_{\overline c} \, {\overline c}} \right] \cdot
\left[ \frac{b_{\overline q} \, {\overline q}}{b_q\, q} \right] 
\label{oper_q}
\end{eqnarray}

The factor ${\cal S}^{sc}$ substitutes a strange quark with a charm
one and the factor ${\cal S}^{cq}$ changes
the charm quark into a light one.
These properties lead to the following identity: 
\begin{equation}
{\cal S}^{qs} \cdot {\cal S}^{sc} \cdot {\cal S}^{cq} \equiv  1
\end{equation}

This identity can be rewritten as an identity between the mesonic ratios:
\begin{equation} 
\frac{{\overline D_s} / D_s}{{\overline D} / D} = 
{\overline K}/{K}
\end{equation}

A comparison of this simple relation with experimental data  
could serve as test of quark matter formation and
coalescence type rehadronization in heavy ion collisions.

The inverse of the step factors is defined as 
${\cal S}^{ji} = ({\cal S}^{ij})^{-1}$.
The  structure of the antibaryon to baryon ratios 
in ALCOR$_c$ is visualized in a geometric manner in Fig. 1.
This way, more complicated but definitely interesting relations can be
obtained. Since the baryons with one charm quark (or antiquark)
can be measured most easily, one may consider the following
relations as candidates for an experimental test:
\begin{equation}
\frac{{\overline \Xi}_c \, ({\overline q} {\overline s} {\overline c})}
{\Xi_c \,(qsc)} 
= {\cal S}^{qs}  \left[\frac{{\overline Y}_c}{Y_c} \right]
= {\cal S}^{qc}  \left[\frac{{\overline Y}}{Y} \right]
= {\cal S}^{sc}  \left[\frac{{\overline \Xi}}{\Xi} \right]
= {\cal S}^{sq}  \left[\frac{{\overline \Omega_c}}{\Omega_c} \right].
\end{equation}
These yield the following simple relation between baryonic and mesonic ratios:
\begin{eqnarray}
\frac{ {\overline Y}/ Y}{{\overline Y}_c / Y_c}
& = & D_s / {\overline D_s} \ ,\\
\frac{ {\overline N}/ N}{{\overline Y}_c / Y_c}
& = & D / {\overline D } \ ,\\
\frac{ {\overline N}/ N}{{\overline Y} / Y}
& = & {\overline K} / K \ .
\end{eqnarray}

\begin{center}
\vspace*{14.0cm}
\includegraphics{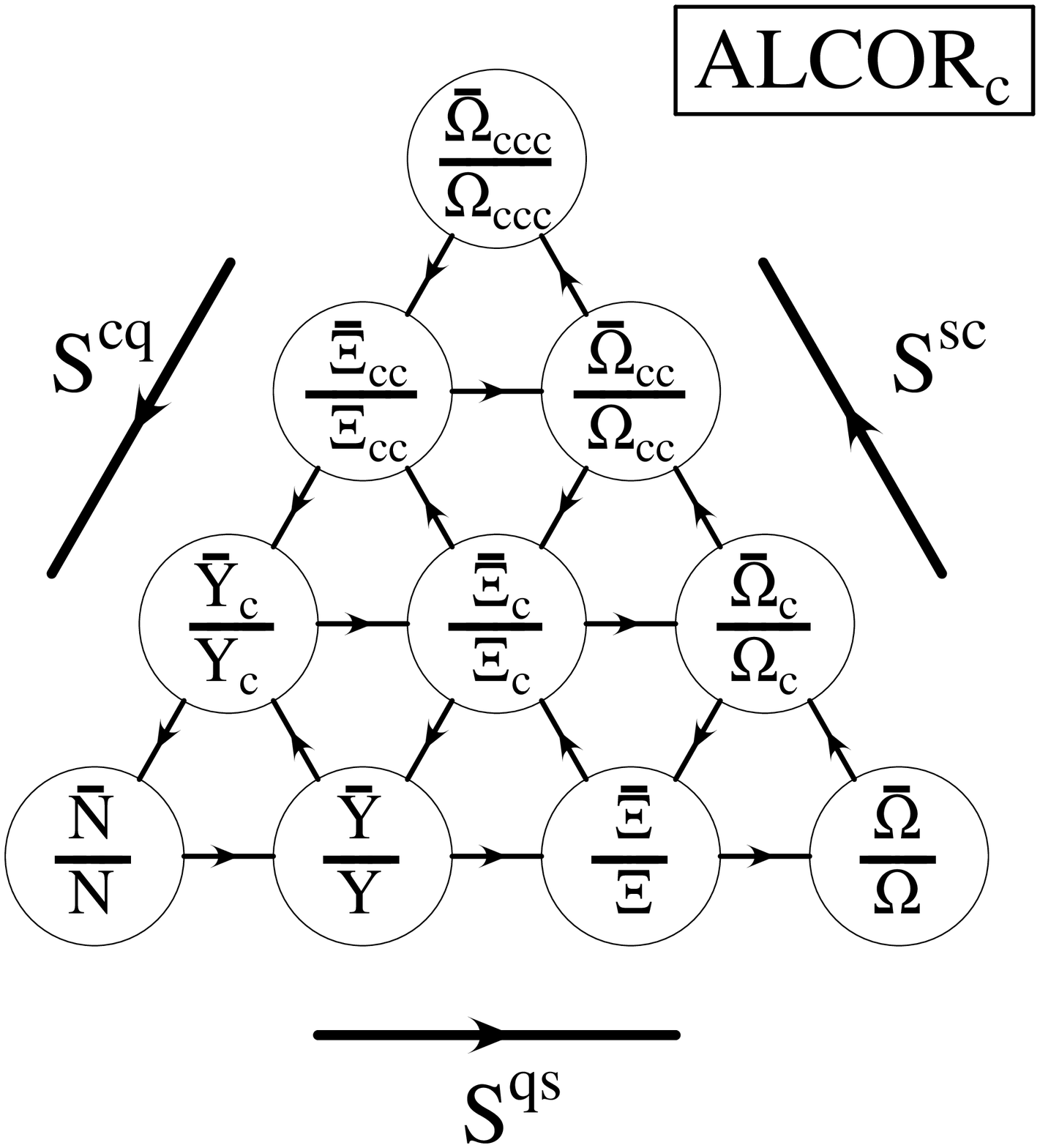}
\vspace*{-1.5 truecm}
\begin{minipage}[t]{15.cm}
{ {\bf Fig.~1.}
The application of mesonic step factors ${\cal S}^{qs}$, ${\cal S}^{sc}$ and
${\cal S}^{cq}$ on the antibaryon to baryon ratios. The arrows are
indicating the three corresponding directions of
shifting.
}
\end{minipage}
\end{center}

A number of similar expressions can be derived from Figure 1, picking up
a given ratio and following all the paths to reach that from its
neighbors.
\medskip

{\it In summary }, we have made simple predictions from
the ALCOR$_c$ model, extending the ALCOR model
of algebraic coalescence and rehadronization of quark matter
to the case when charmed quarks and final state hadrons are present
in a significant number. 
We found that the various  ${\overline M}/M$ mesonic ratios 
connect different ${\overline B}/B$ ratios.
The agreement between the obtained theoretical relations 
and those in the measured data could serve as proof 
or disproof of the formation of
quark matter in heavy ion collisions followed by a fast
hadronization via quark coalescence. The predictions made in this
paper are independent from the detailed values of coalescence coefficients,
we have assumed only their symmetry for charge conjugation. 
The calculations of the absolute numbers of produced particles  from ALCOR$_c$
requires the specification of these coalescence coefficients from
calculations of cross-sections. 
\bigskip

{\bf Acknowledgment:}
This work was supported by the OTKA Grants No. T029158, 
 T025579 and T024094.

\end{document}